# Predicting cell phone adoption metrics using machine learning and satellite imagery

**Edward J. Oughton[1*] and Jatin Mathur[2*]**

[1]College of Science, George Mason University, Fairfax, VA
[2]Department of Computer Science, University of Illinois at Urbana-Champaign, Urbana, USA
*Both authors contributed equally to the manuscript and share joint first authorship

Corresponding authors: Edward J. Oughton (e-mail: eoughton@gmu.edu);
Jatin Mathur (e-mail: jatinm2@illinois.edu)

Corresponding address: GGS, George Mason University, 4400 University Drive, Fairfax, VA

# Abstract

Approximately half of the global population does not have access to the internet, even though digital connectivity can reduce poverty by revolutionizing economic development opportunities. Due to a lack of data, Mobile Network Operators and governments struggle to effectively determine if infrastructure investments are viable, especially in greenfield areas where demand is unknown. This leads to a lack of investment in network infrastructure, resulting in a phenomenon commonly referred to as the 'digital divide'. In this paper we present a machine learning method that uses publicly available satellite imagery to predict telecoms demand metrics, including cell phone adoption and spending on mobile services, and apply the method to Malawi and Ethiopia. Our predictive machine learning approach consistently outperforms baseline models which use population density or nightlight luminosity, with an improvement in data variance prediction of at least 40%. The method is a starting point for developing more sophisticated predictive models of infrastructure demand using machine learning and publicly available satellite imagery. The evidence produced can help to better inform infrastructure investment and policy decisions.

# 1. Introduction

How do we predict local cell phone adoption? And, could the combined purchasing power of households viably attract new digital infrastructure investment, for example, for 4G or 5G?

Currently, digital ecosystem actors such as governments, regulators, and development agencies, as well as many Mobile Network Operators (MNOs), lack this vital insight in unconnected areas. Hence, infrastructure deployment is perceived as riskier in these places, often preventing needed investment, leading to a 'digital divide' between those with voice and data access, and those without. Ultimately, having internet access allows users to participate in the digital economy, providing revolutionary economic and societal opportunities. While those without access are left behind.

The United Nation's Sustainable Development Goals (SDGs) provide a vision for achieving a better future for all which can be sustained over the long term (United Nations, 2019), and SDG target 9.c places a special focus on delivering universal and affordable broadband to help reduce poverty. Over many decades, Information and Communications Technologies (ICTs) have been seen as a key way to enable digitally-led economic development and help deliver the SDGs, potentially lifting millions out of poverty (Haile et al., 2019; Mansell, 2001, 1999; Mansell and Wehn, 1998). Hence, solving the digital divide is critical to this mission. Currently, universal broadband still remains an ambitious goal even in established frontier economies such as the United States, demonstrating the challenge of achieving viable internet economics in rural and remote areas (Claffy and Clark, 2019). Particularly in the current era of the digital economy, users do not just require a stable 2G voice connection, but also a reliable data connection to enable functionality for the existing range of applications and services. 4G is the most common data connectivity technology, but this will increasingly be supplemented by 5G over the next decade (Hidalgo et al., 2020; Oughton et al., 2020; Tchamyou et al., 2019).

The importance of the analysis reported in this paper is highlighted when we consider the vast quantities of capital invested into digital divide projects globally every year. As just one example, the World Bank's Digital Development program aims to provide the necessary knowledge and financing to help close the global digital divide, ensuring countries can take full advantage of the adoption of internet-based technologies for economic development purposes. Over the past four years alone, the World Bank has invested over $3.8 billion (USD) in ICT projects (World Bank, 2019), with over $1.1 billion going to countries on the African continent, as outlined in Table 1.

*Table 1 World Bank ICT financing (World Bank, 2019)*

| Region | 2015 ($m) | 2016 ($m) | 2017 ($m) | 2018 ($m) | 2019 ($m) | Total ($m) |
|---|---|---|---|---|---|---|
| Africa | 159 | 44 | 274 | 226 | 471 | 1,174 |
| East Asia & Pacific | 69 | - | 207 | 80 | 140 | 496 |
| Europe & Central Asia | 42 | 39 | 8 | 129 | 116 | 334 |
| Latin America & Caribbean | 48 | - | 122 | 13 | 46 | 229 |
| Middle East & North Africa | - | 145 | 183 | 62 | 308 | 698 |
| South Asia | 38 | 43 | 228 | 232 | 310 | 851 |
| Annual lending | 356 | 271 | 1,022 | 742 | 1,391 | 3,782 |

With such vast investments targeting the digital divide, there is strong motivation to develop data-driven broadband strategies to inform both telecom investment decisions and policies (Feijóo et al., 2020; Feijóo and Kwon, 2020; Taylor and Schejter, 2013). Estimating the supply-side costs for building new digital infrastructure has been well researched and is a combination of the equipment cost (which does not vary much between countries), and country-specific cost factors relating to geography, labor and taxation (Chiaraviglio et al., 2017; Jha and Saha, 2017; Oughton and Russell, 2020; Ovando et al., 2015). In contrast, on the demand-side, there is considerable spatial heterogeneity in adoption between one local area to the next, dependent on a range of socio-economic, demographic and cultural factors (Blank et al., 2018; Francis et al., 2019; Kabbiri et al., 2018; Oughton et al., 2015; Rhinesmith et al., 2019; Whitacre et al., 2015).

In an ideal situation, potential MNOs, infrastructure investors and policy decision-makers would have a good knowledge of the potential demand in a local area, covering both barriers and pent-up demand to aid business and planning decisions (Martínez-Domínguez and Mora-Rivera, 2020; Mossberger et al., 2012; Owusu-Agyei et al., 2020; Reddick et al., 2020; Rosston and Wallsten, 2020; Taufique et al., 2017). Information would be collected via on-the-ground survey methods and cover (i) existing cell phone ownership for basic, featurephone and smartphone devices, (ii) the current Average Revenue Per User (ARPU), and (iii) the potential willingness to pay for new services. This information could be combined to estimate initial and future potential revenue for MNOs, and the effective magnitude of required state subsidies to make deployment viable. While these surveys can provide rich context for informing digital divide policies, they can be expensive to undertake due to labor-intensive methods. A survey also only provides information for a single point in space.

Often therefore data must be generalized from those few areas with known demand information to unknown locations, introducing uncertainty which can affect network design, private financing, and government support through schemes such as network subsidies. Ideally, we would like a predictive model that can better inform this data generalization process. Thus, there is motivation to explore new analytical options for quantifying the digital divide and providing improved evidence to design policies to reduce digital inequalities. Such evidence is essential for governments and international aid agencies (Maitland et al., 2018).

Currently, there has been much development around the use of machine learning techniques to enhance telecom decision making (Balmer et al., 2020; Righi et al., 2020; Vesnic-Alujevic et al., 2020). However, this poses a significant challenge for digital divide researchers because there are very few critical assessments of the effectiveness of these techniques. Indeed, researchers should not accept machine learning conjecture without independent quantitative assessment of these methods, with such assessments conforming to the highest standards of scientific reproducibility. Considering these issues, a single research question is now identified to address in this paper.

*How effective are different techniques at predicting cell phone adoption metrics from satellite imagery, such as device penetration and monthly spending on telephone services?*

In answering this question, the key contributions of this paper include:

1. Providing a validated method for predicting cell phone adoption metrics from satellite images.
2. Evaluating independent quantitative data on the effectiveness of machine learning techniques, over existing approaches.
3. Developing a documented open-source codebase for the digital divide community to access, reproduce the results and further develop the method, via the Telecom Analytics for Demand using Deep Learning online repository:

Having articulated the main contributions, the structure of the paper is now outlined. Section 2 is a literature review focusing on the digital divide literature, and existing metric prediction from satellite imagery. Section 3 details the method employed, before reporting results in Section 4. Limitations with the method are covered in Section 5. Finally, a discussion is undertaken in Section 6 and conclusions are presented in Section 7.

## 2. Literature review

Two areas of literature are pertinent to the research question, including issues associated with market failure and the digital divide, and existing analyses which have used satellite imagery to predict metrics of interest.

### 2.1. Market failure and the digital divide

Mobile network infrastructure is generally delivered via market-based methods. Indeed, evidence suggests that market competition combined with a suitable regulatory environment is positively correlated with telecom performance and better consumer outcomes, encouraging such an approach (Bauer, 2010; Cave, 2006; Wallsten, 2001),. Hence, investments are generally made based on rational infrastructure investment decisions by profit maximizing private operators. Thus, there must be a viable return which can feasibly be made, for the necessary infrastructure to be deployed.

The problem therefore is that in areas of demand uncertainty (often exacerbated by a lack of data), the necessary infrastructure required for economic development is not delivered leading to market failure (Oughton et al., 2018; Thoung et al., 2016). Although solving coverage issues alone may not eliminate the digital divide (Reisdorf et al., 2020), basic infrastructure is a necessary prerequisite to gaining sustainable economic development benefits from digital technologies and the wider positive societal impacts (Chen et al., 2020; Chester and Allenby, 2019; Farquharson et al., 2018; Graham and Dutton, 2019; Hall et al., 2016; Parker et al., 2014; Saxe and MacAskill, 2019). Thus, market failure issues must be addressed. A variety of technology, business model and policy options are available to attempt to do this and usually focus on using wireless technologies as the costs of deployment are lower. An essential prerequisite however is mutual collaboration between private MNOs and governments, as improvements in coverage and capacity are most effective when there is simultaneous growth in both infrastructure and spectrum portfolios to enabling scale economics (Peha, 2017).

From an economic perspective, reducing the digital divide usually involves subsidizing both investments in rural areas and services for low-income people (Rosston and Wallsten, 2019). While the focus of the digital divide debate is very often on supply-side coverage gaps or connection speed differentials, the roll-out of infrastructure to unviable locations must ultimately be accompanied by demand-side programs to increase device ownership and digital literacy, as these are key determinants of adoption (Hauge and Prieger, 2010). Too often, digital divide issues are heavily compounded by existing socio-economic disparities, meaning lower income groups can be most affected (Riddlesden and Singleton, 2014). This can too often have a greater disproportionate effect on minority ethnic groups (Gant et al., 2010; Turner-Lee and Miller, 2011).

Estimating demand metrics particularly in greenfield areas is a serious challenge for both MNOs (Suryanegara, 2018), telecom regulators and analysts (Oughton et al., 2019a, 2019b), leading to simplified modeling assumptions which do not necessarily reflect reality. Building new infrastructure is a balancing act (Greenstein, 2010), between delivering to areas of guaranteed demand (motivated by profit maximizing behavior), and incrementally rolling out new infrastructure to areas where coverage is needed but take-up of new services is uncertain (motivated by equitable access policies).

Although revenue metrics are frequently developed, they are rarely translated into spatial estimates of how and where infrastructure investment should next be directed, which for unviable areas may require government action (Sevastianov and Vasilyev, 2018; Vincenzi et al., 2019). Similarly, in forecasts of user adoption for cellular technologies (Jahng and Park, 2020; Jha and Saha, 2020; Kalem et al., 2021; Maeng et al., 2020; Neokosmidis et al., 2017), MNOs are left with very little spatial understanding of how many potential users of new services there might be in each local area, despite this being important.

In conclusion, it would be beneficial to have new evidence on local adoption of cell phone metrics to help inform both private and governmental actions to reduce the digital divide.

## 2.2. Metric prediction from satellite imagery

While cell phone adoption has been studied for many countries, including across Africa (Wesolowski et al., 2012), researchers usually focus on analyzing survey data, with few attempt to develop predictions at the national scale. This is surprising given that internet-enabled technologies are increasingly being used to address a range of issues relating to health, climate change, economic development, and disaster resilience. Therefore, it is essential to know who is connected, and where.

Currently there is considerable research which uses cell phone call records or location data, obtained from MNOs, to metrics of interest, such as population density (Deville et al., 2014), urban growth (Bagan and Yamagata, 2015), cellular network anomalies (Sultan et al., 2018) and socio-economic characteristics (Fernando et al., 2018; Koebe, 2020; Schmid et al., 2017). However, the limitations of this approach relate to there being (i) no call data in areas with no coverage, and (ii) privacy issues associated with this type of data, affecting data sharing.

It is increasingly common for statistical frameworks to be developed which take advantage of satellite data to augment official statistics. Many papers have focused on using nightlight luminosity data to assess questions relating to economics (Henderson et al., 2012, 2011), human development (Bruederle and Hodler, 2018), urban extent (Zhou et al., 2015), conservation (Mazor et al., 2013), atmospheric composition (Proville et al., 2017) and measuring the post-disaster impacts of natural hazards (Elliott et al., 2015; Gillespie et al., 2014). While the analysis of mobile phone data is well established (Steenbruggen et al., 2015), new developments are taking advantage of a combination of machine learning with call records and satellite imagery, to address a similar set of questions relating to poverty estimation (Ayush et al., 2020; Jean et al., 2016; Perez et al., 2017; Steele et al., 2017), ecosystem monitoring (Cord et al., 2017), estimating land cover types (Goldblatt et al., 2018) and creating data layers relevant to the Sustainable Development Goals (Boyd et al., 2018; Pokhriyal and Jacques, 2017). However, such approaches have rarely been used to assess the digital divide. Importantly, a key advantage of remote sensing using satellite data is that (i) there is access to an abundant, routinely collected body of data, (ii) has very wide geographic coverage of such data allowing scalability across countries, and (iii) has very high spatial resolution (Donaldson and Storeygard, 2016).

An increasingly used technique is transfer learning, where pretrained models are reapplied to new tasks to help tackle data limitations, such as with survey data (Jean et al., 2016). The goal of transfer learning is to reuse low-level learned aspects of the feature domain, from abundant data such as luminosity images or mobile phone records. High-level specific features can then be learnt for problems with the limited data available, preventing the need to fit a model from scratch. Several types of transfer learning have been surveyed in the literature (Pan and Yang, 2010), but inductive transfer learning is a commonly applied approach, where the domain of two machine learning problems are the same, but the task is different.

## 3. Method

The method contains five steps, starting by introducing the available data and then articulating the data preprocessing steps. The concept of transfer learning and how this approach is used to turn an image into a feature vector is explained. Next, we describe how the feature vector is used to predict the metrics of interest. Lastly, we explain how to generalize the model to new regions.

## 3.1. Available data

To obtain measurements of the metrics of interest, data are taken from the World Bank's Living Standards Measurement Survey (LSMS), a multi-topic household survey undertaken in partnership with various national statistical offices. The survey collects up-to-date information for measuring poverty, livelihood, and living conditions for specific household clusters in space, but is therefore not comprehensive across a whole country due to the prohibitive cost of surveying very large areas. The data are collected at what is called the "cluster" level – a small geographic region with a distinct latitude and longitude. We need to generalize this data to develop a national dataset. Data are downloaded for the Malawian Fourth Integrated Household Survey 2016-2017 (World Bank, 2016a) and Ethiopian Socioeconomic Survey 2015-2016 (World Bank, 2016b), reported by metric in Table 2. Penetration is defined by the percentage of households with at least one cellphone, and consumption of phone services is based on the monthly spending on telephone services per capita (which is broadly similar to the Average Revenue Per User).

*Table 2 LSMS telecom metrics (household level)*

| Country | Variable description | WB Data file name | Column | Households surveyed | Point clusters | Source |
|---|---|---|---|---|---|---|
| Malawi | Household has a phone | hh_mod_f.csv | hh_f34 | 12,447 | 780 | (World Bank, 2016a) |
| Malawi | Spend on phone services | hh_mod_f.csv | hh_f35 | 12,447 | 780 | (World Bank, 2016a) |
| Ethiopia | Household has a phone | sect9_hh_w3.csv | hh_s9q22 | 4,954 | 530 | (World Bank, 2016b) |
| Ethiopia | Spend on phone services | sect9_hh_w3.csv | hh_s9q23 | 4,954 | 530 | (World Bank, 2016b) |

Surveys are conducted in each cluster, and for the sake of anonymity LSMS cluster coordinates are offset by a small random amount. In Malawi, there are 780 clusters and 12,447 households surveyed, whereas in Ethiopia, there are 530 clusters and 4,954 household surveyed. To obtain images, the Planet web Application Programming Interface was used to query daytime *PlanetScope* satellite images in the time range (2014-2016) at a zoom level of 14 (based on 'PSScene3Band' with a ~3m resolution). We always use the latest timestamped image to obtain the closest visit to when the on-the-ground survey was undertaken, to reduce uncertainty in the analysis. The images had a resolution of 256x256 pixels. A cloud cover filter of 5% was applied (removing images with more than 5% cloud coverage). The challenge is to be able to predict the desired metrics of interest from these images.

## 3.2. Data preprocessing

There are two cases to consider: 1) training a model to work in a single country ("single-country"), and 2) generalizing the model to work on multiple countries ("cross-country"). Cross-country generalization is limited in this study as only data from Malawi and Ethiopia is used. However, this analysis is conducted as a baseline to enable future cross-country improvements. Using the LSMS data, 10x10 km bounding box is generated around the geometric centroid of each surveyed cluster. For each bounding box 20 download locations are uniformly samples. For Malawi, 780 clusters with 20 images per cluster leads to 15,600 images, and in Ethiopia, 530 clusters with 20 images per cluster leads to 10,600 images. In accordance with previous work, we bin each metric into four numeric ranges ("bins") identified using a quantile cut (Jean et al., 2016). Each cluster and its images are assigned to the corresponding bin. The Convolutional Neural Network (CNN) is trained to classify an image's bin. Lastly, the clusters that should be held out from the training process are identified to properly validate the model. In the single-country case, 30% of the clusters are randomly held out for validation. In the cross-country case, an entire country is held out. The limitations of this approach are discussed later in the paper.

## 3.3. Transfer learning

Images are a very complex data source. Besides numerous raw inputs, images have many hard-to-quantify factors such as relative position, orientation, and shading making them difficult for machines to start using from scratch. As such it has become common practice to use a technique called transfer learning to "transfer" model learning in one context to another. Specifically, a parameter transfer method is used with the aim of fine-tuning pre-trained models on the data at hand (Houlsby et al., 2019; Kumagai, 2017). In this application, a pretrained University of Oxford Visual Geometry Group (VGG) model is chosen, which is widely used because the architecture is both highly effective and open-source (Kolar et al., 2018; Simonyan and Zisserman, 2015). Using this approach, the VGG model is trained on the ImageNet dataset which contains millions of images and over 1000 subjects (e.g. trees, vehicles, buildings etc.) (ImageNet, 2020). By pretraining on ImageNet, the VGG model is a very good tool for parameter transfer learning, as specifically (i) the domains are the same (images) and (ii) the VGG model has been extensively exposed to different domains and learned to extract useful information. For those not familiar with this method, the approach is metaphorically like a language student learning from a wide variety of materials (books, audio sources, web resources etc.), prior to beginning conversational engagement with a human. The prior step provides a basic structural understanding of a language, whereas the second step helps to refine and further develop the existing understanding to develop fluency. For an introductory overview of transfer learning methods and applications, there are resources available in the literature (Sarkar et al., 2018; Yang et al., 2020).

To train the model, the pretrained VGG network is downloaded via PyTorch (PyTorch, 2020). Specific layers are reinitialized to function on a four-output classification task (as per the binning process described in Section 3.2). An image preprocessor is added during training that randomly chooses a subsection of the image to crop and pass to the CNN, preventing the CNN from being handed the same image repeatedly, thereby reducing overfitting (which is a common problem for deep learning models). A learning rate of $3 \times 10^{-6}$ is used with a batch size of 8, along with a custom loss function and the Adam optimizer. The custom loss function is designed to mitigate the issue of assigning a real-valued variable into bins. For a significant number of cases, clusters will be "close" to the bin boundary, but that information will be lost because of the binning process. Cross-entropy loss aims to maximize the probability of the correct class and reduce error, yet for continuous variables the concept of correct class is artificial. The custom loss function defines anything in the top 10% of a bin to be "close" to the higher bin, and anything in the bottom 10% of a bin to be "close" to the lower bin. For images that are not "close" to another bin, a regular cross-entropy loss is applied. For images that are "close" to another bin, the loss function is the following in equation (1):

$$L(o, l) = \alpha * CEL(o, l) + (1 - a) * CEL(o, N(l)) \qquad (1)$$

Where $o$ is a vector representing the real-valued predictions for each bin and $l$ is the integer label for the correct bin. $CEL$ refers to the Cross-Entropy Loss. $N(l)$ returns the integer label of the nearby bin (the bin the cluster is "close" to). A weighting factor ($a$) assigns a degree of priority to the true class and a degree of priority to the nearby class. This custom loss function prevents the model from being punished too harshly if it predicts the nearby class, in cases where there is ambiguity about which bin the cluster belongs to. The first 5 epochs are used to train only the new layers (all other layers are "frozen" to use PyTorch terminology). Another 25 epochs are spent training the entire model.

## 3.4. Prediction

The vector output of a layer near the end of the CNN is used as a feature vector representation of the image, with each layer being reported in Table 3. After the CNN finishes training, layer 33 is specifically extracted, with an output vector of length 4096. Layer 36 is reinitialized for the four-output class rather than the previous 1000-output class.

*Table 3 Modified PyTorch VGG architecture*

| Layer | Description | Layer | Description | Layer | Description |
|---|---|---|---|---|---|
| 0 | Conv2d | 12 | BatchNorm2d | 24 | ReLU |
| 1 | BatchNorm2d | 13 | ReLU | 25 | Conv2d |
| 2 | ReLU | 14 | MaxPool2d | 26 | BatchNorm2d |
| 3 | MaxPool2d | 15 | Conv2d | 27 | ReLU |
| 4 | Conv2d | 16 | BatchNorm2d | 28 | MaxPool2d |
| 5 | BatchNorm2d | 17 | eLU | 29 | AdaptiveAvgPool2d |
| 6 | ReLU | 18 | Conv2d | 30 | Linear |
| 7 | MaxPool2d | 19 | BatchNorm2d | 31 | ReLU |
| 8 | Conv2d | 20 | ReLU | 32 | Dropout |
| 9 | BatchNorm2d | 21 | MaxPool2d | 33 | Linear |
| 10 | ReLU | 22 | Conv2d | 34 | ReLU |
| 11 | Conv2d | 23 | BatchNorm2d | 35 | Dropout |
| | | | | 36 | Linear |

Image feature vectors are averaged per cluster to find an aggregate cluster feature vector. Using only the clusters reserved for training, both random cross-validation and spatial cross-validation are performed to fit five models, each trained on four-fifths (4 "folds") of the training data and validated on the other fifth (the "fold" held out). Each time a fold is held out, a hyperparameter search is performed internally on the four training folds. The only hyperparameter in Ridge Regression is the regularization coefficient. A list of potential regularization coefficients is enumerated, and for each coefficient an "inner" cross-validation is undertaken on the four folds. The coefficient with the best average $R^2$ is chosen. It is important to remember that this hyperparameter search does not pick the coefficient that worked best on the original 5$^{th}$ fold held out, but rather the one that works best during the "inner" cross-validation. This tests generalization onto the 5$^{th}$ fold correctly.

The five models create an ensemble by applying spatial cross-validation and implementing an equal voting scheme that averages the predictions of the five models. Given the limited number of samples, this captures input variability better. The spatially validated models are then used because the $R^2$ of those models on the training set tends to be far closer to the generalized $R^2$ on the validation set. Finally, the model ensemble is tested on the validation clusters. For the single-country case, the validation clusters are the 30% of clusters held out. For the cross-country case, the validation clusters are all clusters belonging to the country held out.

Prediction intervals are computed using a probabilistic formulation of linear regression, as shown in equation (2):

$$argmax(\sigma, \alpha) \prod P(y_i|x_i) = \frac{1}{(2\pi\sigma^2)^{N/2}} e^{\frac{-\alpha}{2\sigma^2}} \qquad (2)$$

Where $\alpha = ||Y - Xw||^2$. $Y$ is a N x 1 matrix of observed values, and $X$ is a N x 4096 matrix of features. $y_i$ refers to the *i*th row of $Y$, as does $x_i$ with $X$. $w$ is the vector of linear weights. Note that this is equivalent to minimizing the classic linear regression objective. Once $w$ is solved, $\alpha$ becomes a constant, which enables $\sigma$ to be solved as in equation (3):

$$\sigma = \sqrt{\frac{\alpha}{N}} \qquad (3)$$

These equations do not strictly apply to the method for two reasons: 1) the method includes L2 regularization (hence ridge regression) in its objective, and 2) the method creates an ensemble of regression models. For the sake of simplicity, the average of all ridge regression $w's$ will be substituted into $\alpha$. These two simplifications are not too drastic because regularization maintains a very similar objective and an ensemble of linear models is equivalent to a single linear model that contains averages of all model weights. With these simplifications $a$ can be determined (and then $\sigma$).

To compare the results to a baseline, non-CNN models are constructed based on (i) population density, and (ii) nighttime luminosity. Population density is a common way to make disaggregated estimates of telecom demand, for example, in telecom regulatory decision support models which utilize different urban, suburban and rural 'geotype' settlement patterns (Ofcom, 2018). Additionally, nighttime luminosity has also been used to scale telecom demand, for example in estimating the Average Revenue Per User (ARPU) (Oughton, 2021; Oughton et al., 2021; Oughton and Jha, 2021).

Specifically, Ridge Regression is applied using the same cross-validation techniques with a model ensemble and held-out clusters. Nightlight luminosity data are collected using annual composites from 2015 via the Visible Infrared Imaging Radiometer Suite (VIIRS) dataset. Population data obtained from the World Population (WorldPop) 1 km$^2$ raster data layer (Stevens et al., 2015; Tatem, 2017). This data is averaged across the 10x10 km bounding box around each surveyed cluster and used directly to predict the telecoms demand metrics.

### 3.5. Application step

Each country boundary is extracted from the Global Administrative Database of Areas (GADM) (GADM, 2019) and split into 10x10 km grid squares. In total, 20 images are downloaded per grid and passed through the CNN to obtain their feature vector representation. Vectors are averaged across each grid to get a feature vector per grid. This is passed through the ensembled ridge regression model to obtain predictions for each metric per grid tile. Grid squares with very low populations are dropped to avoid these affecting the results. Finally, predictions are mapped. Figure 1 summarizes the method process, from model creation, to prediction validation, to application.

*Figure 1 Method overview*

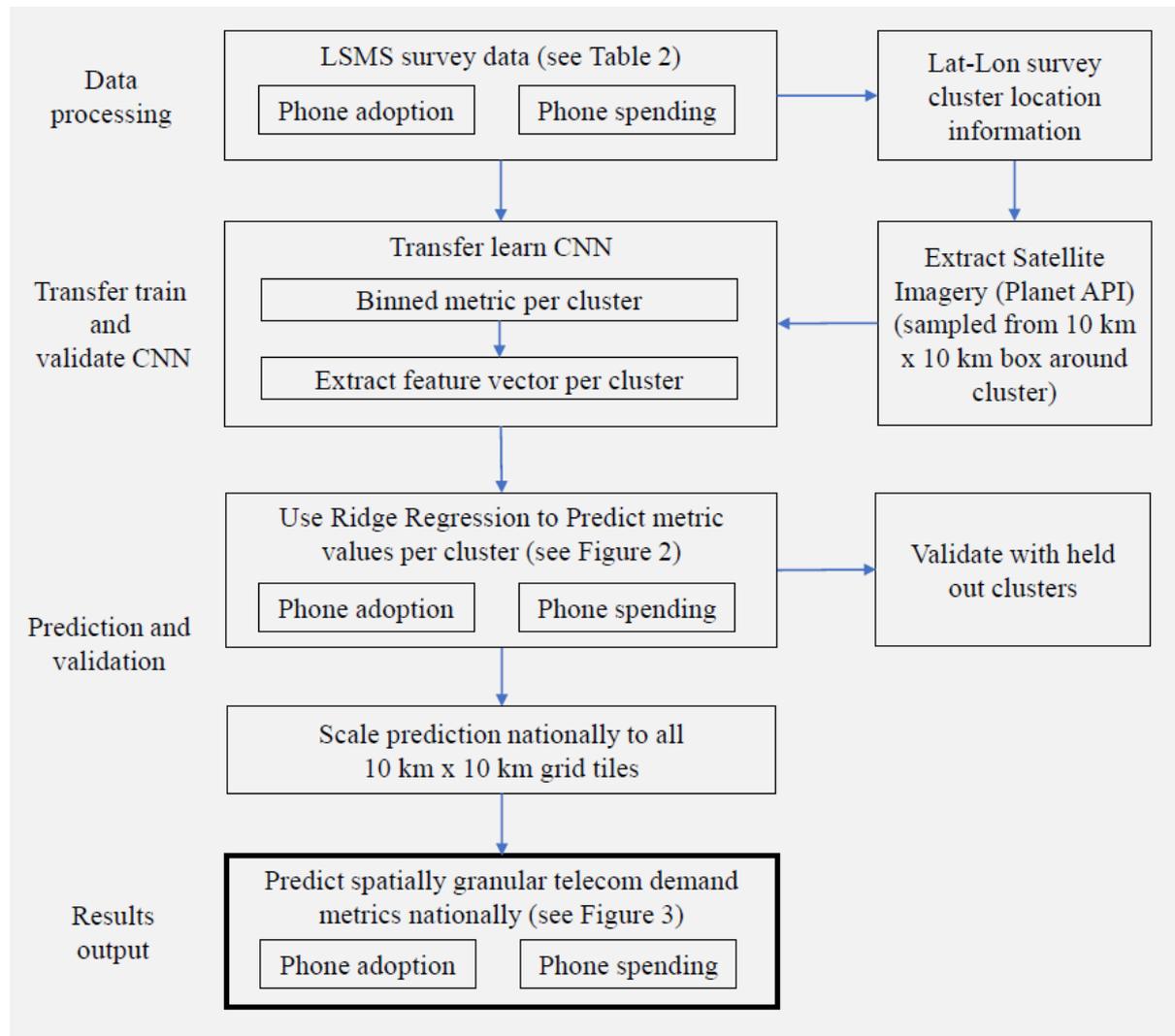

## 3.6. Contextual background to selected countries

Malawi is a landlocked country in Southern Africa, sharing borders with Mozambique, Zambia, and Tanzania. The population is expected to double over the next two decades, from the 19 million citizens present in 2019. As a low-income country, Malawi is one of the poorest in the world with nearly 80% of the population dependent on agriculture making deployment of new digital infrastructure such as 4G challenging, (World Bank, 2021a). The two main mobile operators include

Airtel Malawi Limited and Telekom Networks Malawi Limited, although a third national license has recently been granted. Policy guidance is provided by the Ministry of Information and Communications Technologies, and the Malawi Communications Regulatory Authority (MACRA) is responsible for regulating the sector (International Telecommunication Union, 2018). The spectrum allocation approach by MACRA is based on a "first-come, first-served" basis so long as frequencies are available, with more competitive processes being introduced should there be a spectrum shortage (MACRA, 2021).

The other selected country is Ethiopia which has a strategically important location in the Horn of Africa, and is also landlocked, sharing a border with Eritrea, Somalia, Kenya, South Sudan, and Sudan. Ethiopia has the second largest population in Africa with more than 112 million people in 2019. It has the fastest growing economy in the region, but also one of the lowest per capita incomes (World Bank, 2021b). Previously the sector was dominated by the government-owned Ethio Telecom, however this is now changing (International Telecommunication Union, 2018). The sector is in the process of being privatized with large sums of Ethio Telecom being sold as part of a liberalization agenda, while also issuing new MNO national licenses, with the aim of attracting foreign investment (The Africa Report, 2020). The mobile sector is the responsibility of the newly established Ethiopian Communications Authority (ECA) created in 2019 (Ethiopian Communications Authority, 2019) which intends to allocate spectrum to MNOs efficiently via auction methods (Bloomberg, 2020; Cave and Nicholls, 2017).

Both countries are yet to achieve comprehensive mobile broadband coverage, for example, 4G coverage is at 16% in Malawi and 61% in Ethiopia (GSMA, 2020). Therefore, the proposed satellite imagery-based approach could be incredibly useful to develop new deployment strategies. For example, in Malawi there is quite a substantial proportion of the population needing to be covered (>80%) meaning these analytics could inform future roll-out. Similarly in Ethiopia, with the market liberalization taking place the two new entrants will need to build a substantial amount of greenfield infrastructure, thus requiring analytics to support capital allocation processes.

# 4. Results

The validated CNN accuracies are reported in Table 4, as detailed previously in Section 3.4. As there are four equally represented bins, accuracies above 25% are an improvement over random guessing. Both metrics perform better than the 25% baseline in the single-country case. However, in the cross-country case the accuracies only slightly exceed random guessing.

*Table 4 CNN validation accuracies*

| Type | Name | Binned device penetration | Binned monthly cost |
|---|---|---|---|
| Single-country | Malawi | 44% | 41% |
| | Ethiopia | 39% | 39% |
| Cross-country | Malawi | 29% | 31% |
| | Ethiopia | 30% | 29% |

Validation of the ensembled ridge regression models are shown in Table 5 Model Pearson $R^2$ country validation (best performing models highlighted in yellow) for both single-country and cross-country results. The results indicate that single-country CNN models outperform generalized cross-country usage, and that for both countries and metrics, single-country CNN models far outperform the baseline models.

*Table 5 Model Pearson $R^2$ country validation (best performing models highlighted in yellow)*

| Model | Type | Name | Cross-validation technique | Device Penetration | Monthly Cost |
|---|---|---|---|---|---|
| Population density | Single-country | Malawi | Random | 0.182 | 0.201 |
| | | | Spatial | 0.182 | 0.201 |
| | | Ethiopia | Random | 0.069 | 0.086 |
| | | | Spatial | 0.069 | 0.086 |
| Nightlight luminosity | Single-country | Malawi | Random | 0.211 | 0.183 |
| | | | Spatial | 0.211 | 0.183 |
| | | Ethiopia | Random | 0.083 | 0.152 |
| | | | Spatial | 0.083 | 0.152 |
| CNN | Single-country | Malawi | Random | 0.414 | 0.282 |
| | | | Spatial | 0.410 | 0.284 |
| | | Ethiopia | Random | 0.268 | 0.268 |
| | | | Spatial | 0.268 | 0.268 |
| | Cross-country | Malawi | Random | 0.165 | 0.090 |
| | | | Spatial | 0.169 | 0.124 |
| | | Ethiopia | Random | 0.144 | 0.168 |

|  |  |  | **Spatial** | 0.151 | 0.162 |

The observed versus the predicted values are illustrated in Figure 2 with the associated prediction intervals. Malawi performed much better in predicting cell phone adoption, but only marginally better than Ethiopia in estimating the cost of phone services.

*Figure 2 Observed versus predicted values by metric*

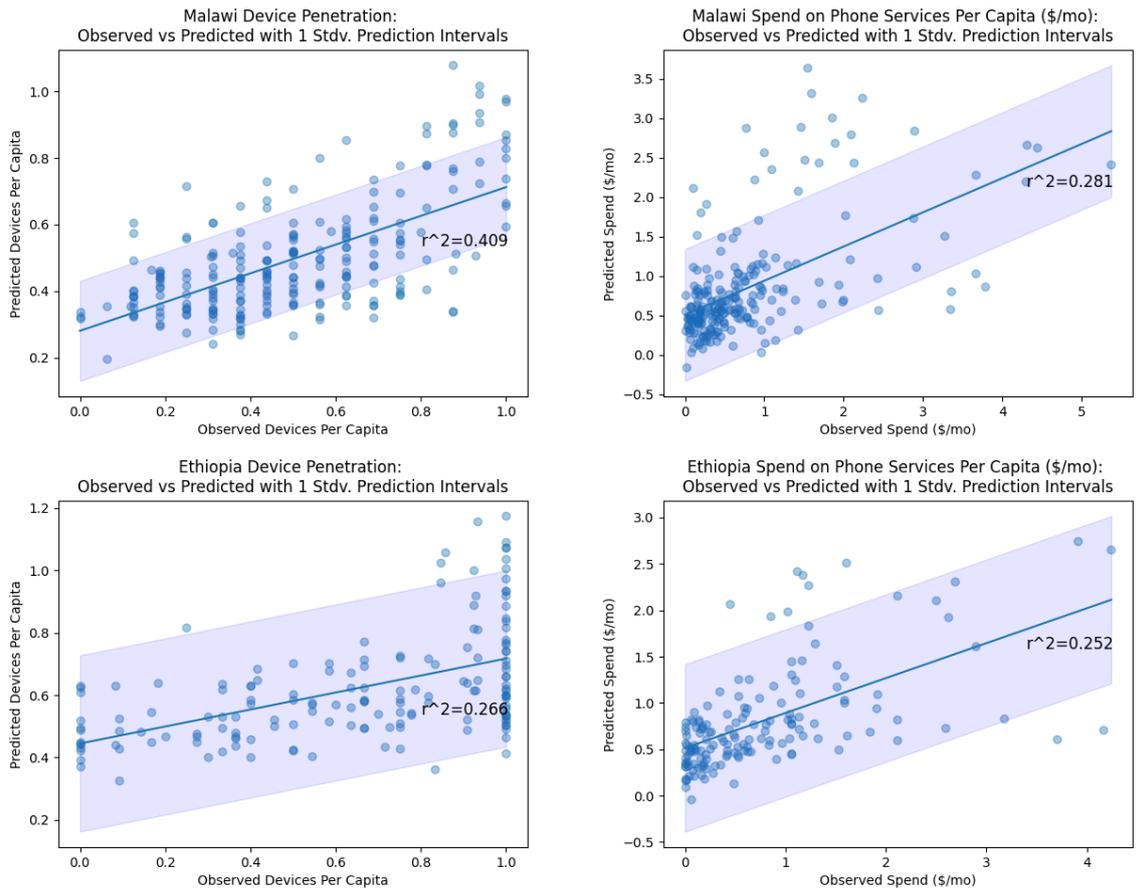

Finally, prediction maps can be created from the results, as shown for Malawi in

Figure 3 at 10x10 km spatial resolution. Evaluation of the spatial results are consistent with expectations. For example, the model estimates higher phone density in the capital Lilongwe in the mid-west area of Malawi, as well as in other populated areas, such as Blantyre in the south east and Mzuzu in the north.

*Figure 3 Predicted device penetration (Left); Predicted monthly cost (Right)*

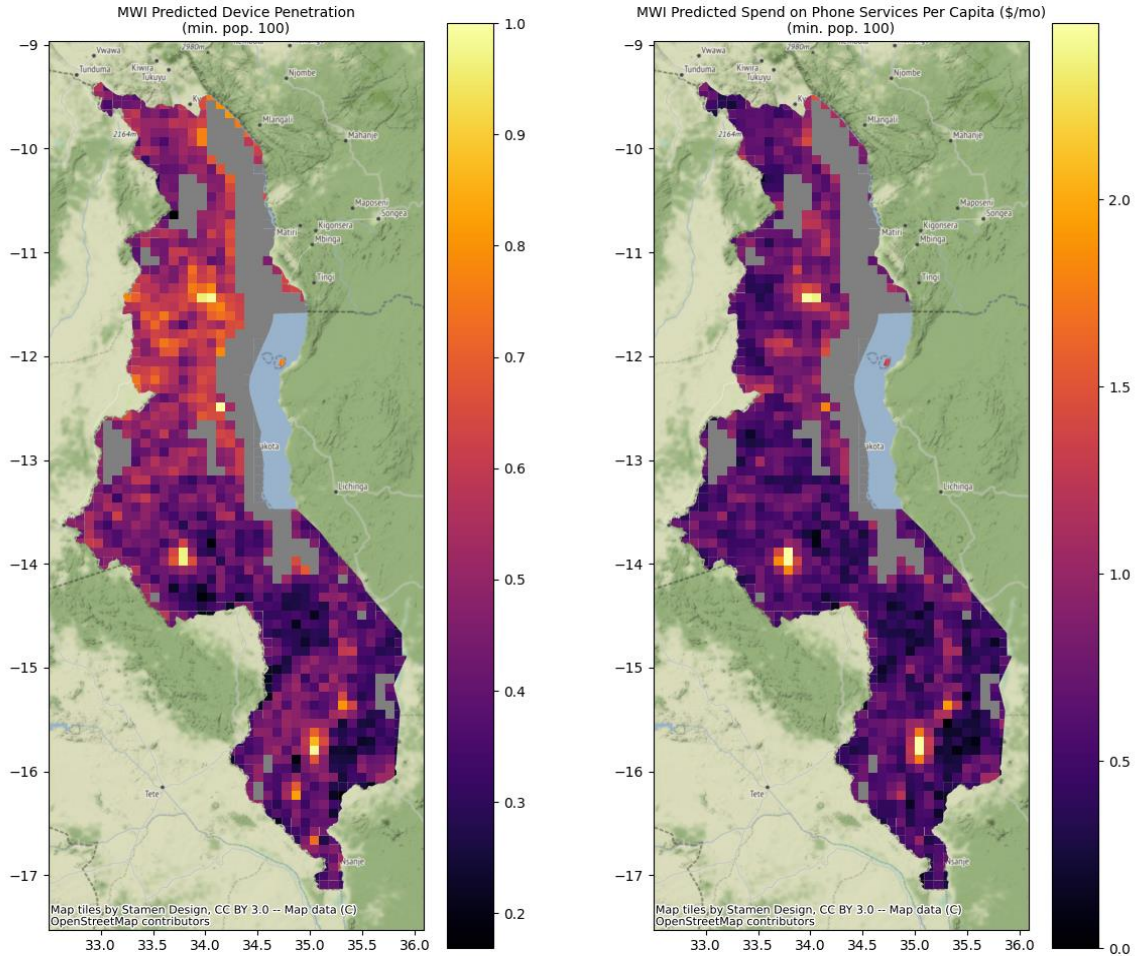

## 5. Limitations and future areas of development

There are three main limitations with the method, including (i) justifying results, (ii) the validation technique, and (iii) satellite image availability. Each of these form future areas of development.

It is quite challenging to explain how a CNN arrives at a predicted result due to the vast quantity of parameters, and the size and structure of the network. This is a widely known issue, raising calls for more explainable machine learning approaches in telecoms (Guo, 2020). Broadly speaking, lack of model interpretability has become an important inhibitor of widespread adaptation of deep learning methods. Activation maps have emerged as one way to interpret a CNN. The idea is simple – project areas that are "activated" by the CNN back onto the original image. This way, it is possible to visually inspect which parts of the image the CNN focuses on. By using activation maps we can evaluate the behavior and limitations of the CNN. One approach to create activation maps is guided backpropagation. This method passes an image through the CNN and performs backpropagation on the known target class.

At the first layer, we store the gradient with respect to each pixel in the image and remove those less than 0. Then, we plot a grayscale map of those gradients. They are the same shape as the original image, and the intensity corresponds to a stronger (positive) gradient. The reason we use a gradient approach is because backpropagation finds partial derivatives; a larger partial derivative can be thought of as a larger contribution.

In

Figure 4 activation maps are presented for three images, with the original satellite image on the left-hand side, and the activation map on the right-hand side. In the top and middle images, we can see that the road network is able to be identified by the CNN. However, large bodies of water can produce strange results, an example being the top left image, where a lake leads to activation of the CNN. Furthermore, in the bottom figure, where the image provider has accidentally inserted a nighttime image, the CNN is also activating. These activations occur in meaningless parts of the image and are much more prevalent than those in either of the two "good" images. As these observations are qualitative and require individual analysis, it is challenging to make a statement about the whole dataset of >20,000 images. Consequently, it is difficult to make a claim about the robustness and generalizability of the CNN itself beyond traditional model validation. However, this means that during application, there could be unpredictable, unexpected, and unexplainable behavior.

*Figure 4 CNN activation maps (Actual images in the left column, and activation maps on the right)*

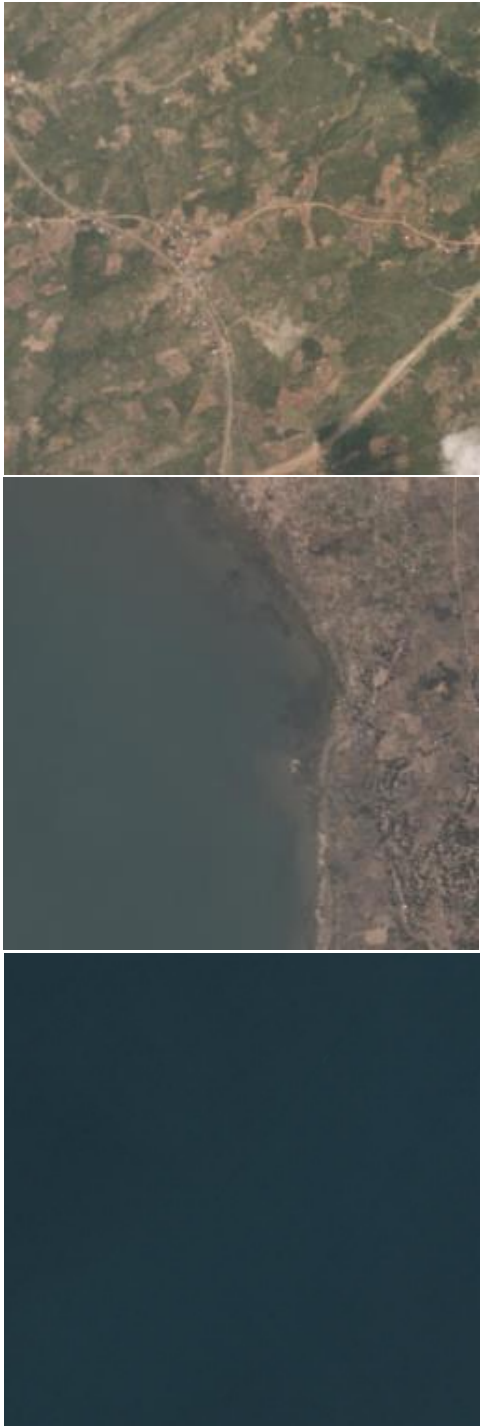
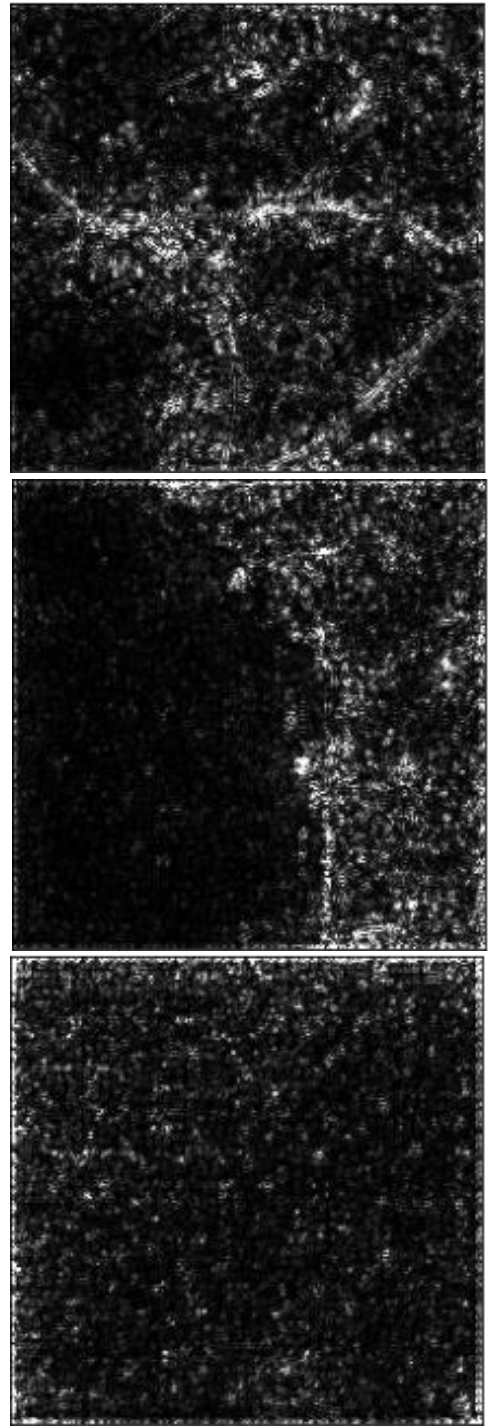

As for model validation, the two biggest constraints are the limited sample size and the long training time. In theory, 5-fold cross-validation would be a good way to test generalization. However, two countries, two metrics, and five folds would mean training a CNN 20 times. Furthermore, instead of randomized cross-validation there is also spatial cross-validation which would make each fold contain clusters that are geographically close. Consequently, each iteration of cross-validation would be testing generalization not only onto unseen clusters, but unseen geographic areas. This prevents the model from training on one cluster and validating on another nearby cluster (which likely has very similar metrics). The advantage of spatial cross-validation is that it more closely reflects the use case of applying the model to new regions. However, this would raise the number of CNN runs to 40. To maintain a reasonable number of training runs, a simple random 30% is held out for validation of single-country models. The downside of this approach is that we do not get the numerical stability that comes with random cross-validation (averaging five results is much better than doing just one), and generalization onto new areas is not tested as thoroughly as possible with spatial cross-validation. Lastly, because the method only involves two countries, further research needs to assess (i) how well the approach works when scaled to many countries, and (ii) what kind of training procedure and data quantity will be sufficient for cross-country generalization.

Finally, a limitation of this type of research relates to obtaining comprehensive and consistent satellite imagery. For the survey years used in this analysis (2015-2017), the field of earth observation was still relatively limited in its ability to produce consistent imagery for whole countries in a single year. This is because obtaining high-quality images is dependent on the revisit rate by satellites in a constellation, as well as the presence of obstructions such as clouds. This meant that in this analysis, the most recent tile over the three-year period (2014-2016) had to be selected, introducing an element of uncertainty. While this is a common limitation of the whole field of earth observation, and not just specific to this study, there are promising developments which will help overcome this issue in the future.

For example, in recent years there has been a rapid increase in the number of commercial Low Earth Orbit observation satellites. With the revisit rates to each tile location rising, this increases the probability of a high-quality cloud-free image being obtained within a defined time-period, helping to boost the temporal resolution of available satellite imagery. Importantly, a key strength of this paper is that the whole codebase has been made open-source and available for the research community to access. Therefore, as we gain improvements in the imagery available, further development of the codebase can take place, refining the results generated here based on more consistent image data.

## 6. Discussion

This section returns the focus to the research question stated in the introduction:

*How effective are different techniques at predicting cell phone adoption metrics from satellite imagery, such as device penetration and monthly spending on telephone services?*

This paper demonstrated a machine learning approach for predicting spatially granular estimates for cell phone adoption with significant improvement over baseline modeling techniques. For example, population density is a common baseline model for predicting cell phone adoption metrics yet only captures approximately 7-20% of the variance in the data across the countries assessed (Malawi and Ethiopia). The use of nightlight luminosity was similar in capturing approximately 8-21% of the data variance. By contrast, the CNN method described in this paper captured up to 41% of the data variance, providing a minimum improvement against the baseline models of at least 40%.

There are several key use cases for the high-resolution, accurate predictions generated by the method in the paper, primarily relating to national assessments. Firstly, international development institutions can quantitatively identify underserved areas and more effectively design interventions for the billions of dollars they invest annually into digital development projects each year in support of the SDGs. Secondly, national and local governments, including telecommunication regulators, can access data which support policy decision making on the digital divide. A standard decision-making tool used in telecoms is the Long Run Incremental Cost (LRIC) model, which is usually spreadsheet-based and focused on modeling a 'hypothetical operator' with average characteristics (e.g. assets, spectrum portfolio, market share etc.). Many assumptions are often used in this approach, particularly relating to the number of cell phones in rural areas and the level of existing demand. Rather than using hypothetical data and assumptions, the method produced here can help to reduce this uncertainty, helping make more effective decisions.

Mobile Network Operators can gain an understanding of two key demand metrics (device penetration and cost per month, which can be also be thought of as average revenue per user) in green field areas where demand is unknown. One key advantage of this method is that it only utilizes satellite images which are globally available. Therefore, the method can be used in data limited locations where we have no cellphone records or electricity usage data. By solely using images, the method learns to associate levels of local development, with the availability of devices and the available consumer purchasing power to spend on phone services. Consequently, in an underserved area with little existing coverage, the method extrapolates the acquired understanding of the relationship between development and telecoms demand onto the new area. Thus, in an application context the approach can predict the general *capability* of residents to own mobile devices and pay for telecom services, based only on widely-available visual evidence.

While there is significant technical complexity to the application of machine learning methods, such as the one presented in this paper, we remain optimistic about their use and application. Firstly, international development organizations are already experimenting with such techniques, meaning this knowledge can be shared with local stakeholders. Secondly, as many national MNOs are part of multinational telecom enterprises, with centralized global strategy and intelligence functions, specialized skills can be developed and applied across many countries. For example, Sonatel in Senegal benefits from strategy and network intelligence input from the company owner Orange, much like Telefonica provides to its myriad South American MNOs, such as Movistar, Peru. When this is combined with the fact that machine learning techniques have become a core part of science and engineering programs at universities around the world, it is increasingly becoming easier for companies to obtain this specialized labor and utilize it to its full advantage (even in resource-constrained economies).

## 7. Conclusion

This paper assessed the effectiveness of different modeling methods at estimating cell phone metrics, such as phone adoption and the capability to pay for cellular services. We find that the baseline models using population density and nightlight luminosity, capture up to 20% and 21% of the data variance respectively. Comparatively, our CNN machine learning approach captured up to 41% of data variance, demonstrating a minimum predictive improvement against the baseline models of at least 40% in all circumstances.

The key contributions of the paper were threefold. Firstly, an accurate and validated method was provided that predicts telecoms demand metrics from satellite images. Secondly, there was a quantitative comparison of this method to existing methods. Finally, the codebase used for the analysis has been made open-source for other researchers and analysts to utilize, reproduce the results, and further develop the method via the online repository: Telecom Analytics for Demand using Deep Learning. Future research needs to be undertaken to expand the assessment method to include other indicators necessary for achieving the SDGs and explore the application of the method to additional countries, including in high-income nations such as the United States.

## References

Ayush, K., Uzkent, B., Burke, M., Lobell, D., Ermon, S., 2020. Generating Interpretable Poverty Maps using Object Detection in Satellite Images. arXiv:2002.01612 [cs].

Bagan, H., Yamagata, Y., 2015. Analysis of urban growth and estimating population density using satellite images of nighttime lights and land-use and population data. GIScience & Remote Sensing 52, 765–780. https://doi.org/10.1080/15481603.2015.1072400

Balmer, R.E., Levin, S.L., Schmidt, S., 2020. Artificial Intelligence Applications in Telecommunications and other network industries. Telecommunications Policy, Artificial intelligence, economy and society 44, 101977. https://doi.org/10.1016/j.telpol.2020.101977

Bauer, J.M., 2010. Regulation, public policy, and investment in communications infrastructure. Telecommunications Policy, Balancing Competition and Regulation 34, 65–79. https://doi.org/10.1016/j.telpol.2009.11.011

Blank, G., Graham, M., Calvino, C., 2018. Local Geographies of Digital Inequality. Social Science Computer Review 36, 82–102. https://doi.org/10.1177/0894439317693332

Bloomberg, 2020. Ethiopia Telecom Auction Set for 2021 With Orange in Contention. Bloomberg.com.

Boyd, D.S., Jackson, B., Wardlaw, J., Foody, G.M., Marsh, S., Bales, K., 2018. Slavery from Space: Demonstrating the role for satellite remote sensing to inform evidence-based action related to UN SDG number 8. ISPRS Journal of Photogrammetry and Remote Sensing 142, 380–388. https://doi.org/10.1016/j.isprsjprs.2018.02.012

Bruederle, A., Hodler, R., 2018. Nighttime lights as a proxy for human development at the local level. PloS one 13.

Cave, M., 2006. Encouraging infrastructure competition via the ladder of investment. Telecommunications Policy 30, 223–237. https://doi.org/10.1016/j.telpol.2005.09.001

Cave, M., Nicholls, R., 2017. The use of spectrum auctions to attain multiple objectives: Policy implications. Telecommunications Policy, Optimising Spectrum Use 41, 367–378. https://doi.org/10.1016/j.telpol.2016.12.010

Chen, P., Oughton, E.J., Tyler, P., Jia, M., Zagdanski, J., 2020. Evaluating the impact of next generation broadband on local business creation. arXiv:2010.14113 [econ, q-fin].

Chester, M.V., Allenby, B., 2019. Toward adaptive infrastructure: flexibility and agility in a non-stationarity age. Sustainable and Resilient Infrastructure 4, 173–191. https://doi.org/10.1080/23789689.2017.1416846

Chiaraviglio, L., Blefari-Melazzi, N., Liu, W., Gutierrez, J.A., van de Beek, J., Birke, R., Chen, L., Idzikowski, F., Kilper, D., Monti, P., Bagula, A., Wu, J., 2017. Bringing 5G into Rural and Low-Income Areas: Is It Feasible? IEEE Communications Standards Magazine 1, 50–57. https://doi.org/10.1109/MCOMSTD.2017.1700023

Claffy, K., Clark, D., 2019. Workshop on Internet Economics (WIE2018) Final Report. ACM SIGCOMM Computer Communication Review 49, 25–30.

Cord, A.F., Brauman, K.A., Chaplin-Kramer, R., Huth, A., Ziv, G., Seppelt, R., 2017. Priorities to Advance Monitoring of Ecosystem Services Using Earth Observation. Trends in Ecology & Evolution 32, 416–428. https://doi.org/10.1016/j.tree.2017.03.003

Deville, P., Linard, C., Martin, S., Gilbert, M., Stevens, F.R., Gaughan, A.E., Blondel, V.D., Tatem, A.J., 2014. Dynamic population mapping using mobile phone data. PNAS 111, 15888–15893. https://doi.org/10.1073/pnas.1408439111

Donaldson, D., Storeygard, A., 2016. The View from Above: Applications of Satellite Data in Economics. Journal of Economic Perspectives 30, 171–198. https://doi.org/10.1257/jep.30.4.171

Elliott, R.J.R., Strobl, E., Sun, P., 2015. The local impact of typhoons on economic activity in China: A view from outer space. Journal of Urban Economics 88, 50–66. https://doi.org/10.1016/j.jue.2015.05.001

Ethiopian Communications Authority, 2019. Our mission [WWW Document]. Ethiopian Communications Authority. URL https://eca.et/our-mission/ (accessed 1.8.21).

Farquharson, D., Jaramillo, P., Samaras, C., 2018. Sustainability implications of electricity outages in sub-Saharan Africa. Nature Sustainability 1, 589–597. https://doi.org/10.1038/s41893-018-0151-8

Feijóo, C., Kwon, Y., 2020. AI impacts on economy and society: Latest developments, open issues and new policy measures. Telecommunications Policy, Artificial intelligence, economy and society 44, 101987. https://doi.org/10.1016/j.telpol.2020.101987

Feijóo, C., Kwon, Y., Bauer, J.M., Bohlin, E., Howell, B., Jain, R., Potgieter, P., Vu, K., Whalley, J., Xia, J., 2020. Harnessing artificial intelligence (AI) to increase wellbeing for all: The case for a new technology diplomacy. Telecommunications Policy, Artificial intelligence, economy and society 44, 101988. https://doi.org/10.1016/j.telpol.2020.101988

Fernando, L., Surendra, A., Lokanathan, S., Gomez, T., 2018. Predicting population-level socio-economic characteristics using Call Detail Records (CDRs) in Sri Lanka, in: Proceedings of the Fourth International Workshop on Data Science for Macro-Modeling with Financial and Economic Datasets, DSMM'18. Association for Computing Machinery, Houston, TX, USA, pp. 1–12. https://doi.org/10.1145/3220547.3220549

Francis, J., Ball, C., Kadylak, T., Cotten, S.R., 2019. Aging in the Digital Age: Conceptualizing Technology Adoption and Digital Inequalities, in: Neves, B.B., Vetere, F. (Eds.), Ageing and Digital Technology: Designing and Evaluating Emerging Technologies for Older Adults. Springer, Singapore, pp. 35–49. https://doi.org/10.1007/978-981-13-3693-5_3

GADM, 2019. Global Administrative Areas Database (Version 3.6) [WWW Document]. URL https://gadm.org/ (accessed 7.11.19).

Gant, J.P., Turner-Lee, N.E., Li, Y., Miller, J.S., 2010. National minority broadband adoption: Comparative trends in adoption, acceptance and use. Joint Center for Political and Economic Studies, Washington D.C.

Gillespie, T.W., Frankenberg, E., Chum, K.F., Thomas, D., 2014. Night-time lights time series of tsunami damage, recovery, and economic metrics in Sumatra, Indonesia. Remote Sensing Letters 5, 286–294. https://doi.org/10.1080/2150704X.2014.900205

Goldblatt, R., Stuhlmacher, M.F., Tellman, B., Clinton, N., Hanson, G., Georgescu, M., Wang, C., Serrano-Candela, F., Khandelwal, A.K., Cheng, W.-H., Balling, R.C., 2018. Using Landsat and nighttime lights for supervised pixel-based image classification of urban land cover. Remote Sensing of Environment 205, 253–275. https://doi.org/10.1016/j.rse.2017.11.026

Graham, M., Dutton, W.H., 2019. Society and the Internet: How Networks of Information and Communication are Changing Our Lives. Oxford University Press.

Greenstein, S., 2010. Building Broadband Ahead of Digital Demand. IEEE Micro 30, 6–8. https://doi.org/10.1109/MM.2010.111

GSMA, 2020. GSMA Intelligence Global Data [WWW Document]. URL https://www.gsmaintelligence.com/ (accessed 2.5.20).

Guo, W., 2020. Explainable Artificial Intelligence for 6G: Improving Trust between Human and Machine. IEEE Communications Magazine 58, 39–45. https://doi.org/10.1109/MCOM.001.2000050

Haile, M.G., Wossen, T., Kalkuhl, M., 2019. Access to information, price expectations and welfare: The role of mobile phone adoption in Ethiopia. Technological Forecasting and Social Change 145, 82–92. https://doi.org/10.1016/j.techfore.2019.04.017

Hall, J.W., Thacker, S., Ives, M.C., Cao, Y., Chaudry, M., Blainey, S.P., Oughton, E.J., 2016. Strategic analysis of the future of national infrastructure. Proceedings of the Institution of Civil Engineers - Civil Engineering 1–9. https://doi.org/10.1680/jcien.16.00018

Hauge, J.A., Prieger, J.E., 2010. Demand-Side Programs to Stimulate Adoption of Broadband: What Works? Review of Network Economics 9. https://doi.org/10.2202/1446-9022.1234

Henderson, J.V., Storeygard, A., Weil, D.N., 2012. Measuring Economic Growth from Outer Space. American Economic Review 102, 994–1028. https://doi.org/10.1257/aer.102.2.994

Henderson, V., Storeygard, A., Weil, D.N., 2011. A Bright Idea for Measuring Economic Growth. American Economic Review 101, 194–199. https://doi.org/10.1257/aer.101.3.194

Hidalgo, A., Gabaly, S., Morales-Alonso, G., Urueña, A., 2020. The digital divide in light of sustainable development: An approach through advanced machine learning techniques. Technological Forecasting and Social Change 150, 119754. https://doi.org/10.1016/j.techfore.2019.119754

Houlsby, N., Giurgiu, A., Jastrzebski, S., Morrone, B., de Laroussilhe, Q., Gesmundo, A., Attariyan, M., Gelly, S., 2019. Parameter-Efficient Transfer Learning for NLP. arXiv:1902.00751 [cs, stat].

ImageNet, 2020. ImageNet Database [WWW Document]. URL http://image-net.org/index (accessed 2.21.21).

International Telecommunication Union, 2018. Measuring the Information Society Report 2018 [WWW Document]. URL https://www.itu.int/en/ITU-D/Statistics/Pages/publications/misr2018.aspx (accessed 1.8.21).

Jahng, J.H., Park, S.K., 2020. Simulation-based prediction for 5G mobile adoption. ICT Express 6, 109–112. https://doi.org/10.1016/j.icte.2019.10.002

Jean, N., Burke, M., Xie, M., Davis, W.M., Lobell, D.B., Ermon, S., 2016. Combining satellite imagery and machine learning to predict poverty. Science 353, 790–794.

Jha, A., Saha, D., 2020. "Forecasting and analysing the characteristics of 3G and 4G mobile broadband diffusion in India: A comparative evaluation of Bass, Norton-Bass, Gompertz, and logistic growth models." Technological Forecasting and Social Change 152, 119885. https://doi.org/10.1016/j.techfore.2019.119885

Jha, A., Saha, D., 2017. Techno-Economics Behind Provisioning 4G LTE Mobile Services over Sub 1 GHz Frequency Bands, in: Sastry, N., Chakraborty, S. (Eds.), Communication Systems and Networks, Lecture Notes in Computer Science. Springer International Publishing, Cham, pp. 284–306. https://doi.org/10.1007/978-3-319-67235-9_17

Kabbiri, R., Dora, M., Kumar, V., Elepu, G., Gellynck, X., 2018. Mobile phone adoption in agri-food sector: Are farmers in Sub-Saharan Africa connected? Technological Forecasting and Social Change 131, 253–261. https://doi.org/10.1016/j.techfore.2017.12.010

Kalem, G., Vayvay, O., Sennaroglu, B., Tozan, H., 2021. Technology Forecasting in the Mobile Telecommunication Industry: A Case Study Towards the 5G Era. Engineering Management Journal 33, 15–29. https://doi.org/10.1080/10429247.2020.1764833

Koebe, T., 2020. Better coverage, better outcomes? Mapping mobile network data to official statistics using satellite imagery and radio propagation modelling. arXiv:2002.11618 [cs, stat].

Kolar, Z., Chen, H., Luo, X., 2018. Transfer learning and deep convolutional neural networks for safety guardrail detection in 2D images. Automation in Construction 89, 58–70. https://doi.org/10.1016/j.autcon.2018.01.003

Kumagai, W., 2017. Learning Bound for Parameter Transfer Learning. arXiv:1610.08696 [cs, stat].

MACRA, 2021. Telecommunications - Spectrum. MACRA. URL https://www.macra.org.mw/?page_id=10995 (accessed 1.8.21).

Maeng, K., Kim, J., Shin, J., 2020. Demand forecasting for the 5G service market considering consumer preference and purchase delay behavior. Telematics and Informatics 47, 101327. https://doi.org/10.1016/j.tele.2019.101327

Maitland, C., Caneba, R., Schmitt, P., Koutsky, T., 2018. A Cellular Network Radio Access Performance Measurement System: Results from a Ugandan Refugee Settlements Field Trial (SSRN Scholarly Paper No. ID 3141865). Social Science Research Network, Rochester, NY.

Mansell, R., 2001. Digital opportunities and the missing link for developing countries. Oxford Review of Economic Policy 17, 282–295.

Mansell, R., 1999. Information and communication technologies for development: assessing the potential and the risks. Telecommunications policy 23, 35–50.

Mansell, R., Wehn, U., 1998. Knowledge societies: Information technology for sustainable development. Oxford University Press.

Martínez-Domínguez, M., Mora-Rivera, J., 2020. Internet adoption and usage patterns in rural Mexico. Technology in Society 60, 101226. https://doi.org/10.1016/j.techsoc.2019.101226

Mazor, T., Levin, N., Possingham, H.P., Levy, Y., Rocchini, D., Richardson, A.J., Kark, S., 2013. Can satellite-based night lights be used for conservation? The case of nesting sea turtles in the Mediterranean. Biological Conservation 159, 63–72. https://doi.org/10.1016/j.biocon.2012.11.004

Mossberger, K., Tolbert, C.J., Bowen, D., Jimenez, B., 2012. Unraveling different barriers to Internet use urban residents and neighborhood effects. Urban Affairs Review 48, 771–810. https://doi.org/10.1177/1078087412453713

Neokosmidis, I., Rokkas, T., Parker, M.C., Koczian, G., Walker, S.D., Siddiqui, M.S., Escalona, E., 2017. Assessment of socio-techno-economic factors affecting the market adoption and evolution of 5G networks: Evidence from the 5G-PPP CHARISMA project. Telematics and Informatics 34, 572–589. https://doi.org/10.1016/j.tele.2016.11.007

Ofcom, 2018. Mobile call termination market review 2018-21: Final statement – Annexes 1 - 15. Ofcom, London.

Oughton, E., 2021. Policy options for digital infrastructure strategies: A simulation model for broadband universal service in Africa. arXiv:2102.03561 [cs, econ, q-fin].

Oughton, E., Tyler, P., Alderson, D., 2015. Who's Superconnected and Who's Not? Investment in the UK's Information and Communication Technologies (ICT) Infrastructure. Infrastructure Complexity 2, 6. https://doi.org/10.1186/s40551-015-0006-7

Oughton, E.J., Comini, N., Foster, V., Hall, J.W., 2021. Policy choices can help keep 4G and 5G universal broadband affordable. arXiv:2101.07820 [cs, econ, q-fin].
Oughton, E.J., Frias, Z., Dohler, M., Whalley, J., Sicker, D., Hall, J.W., Crowcroft, J., Cleevely, D.D., 2018. The strategic national infrastructure assessment of digital communications. Digital Policy, Regulation and Governance 20, 197–210. https://doi.org/10.1108/DPRG-02-2018-0004
Oughton, E.J., Frias, Z., van der Gaast, S., van der Berg, R., 2019a. Assessing the capacity, coverage and cost of 5G infrastructure strategies: Analysis of the Netherlands. Telematics and Informatics 37, 50–69. https://doi.org/10.1016/j.tele.2019.01.003
Oughton, E.J., Jha, A., 2021. Supportive 5G infrastructure policies are essential for universal 6G: Evidence from an open-source techno-economic simulation model using remote sensing. arXiv:2102.08086 [cs, econ, q-fin].
Oughton, E.J., Katsaros, K., Entezami, F., Kaleshi, D., Crowcroft, J., 2019b. An Open-Source Techno-Economic Assessment Framework for 5G Deployment. IEEE Access 7, 155930–155940. https://doi.org/10.1109/ACCESS.2019.2949460
Oughton, E.J., Lehr, W., Katsaros, K., Selinis, I., Bubley, D., Kusuma, J., 2020. Revisiting Wireless Internet Connectivity: 5G vs Wi-Fi 6. arXiv:2010.11601 [cs].
Oughton, E.J., Russell, T., 2020. The importance of spatio-temporal infrastructure assessment: Evidence for 5G from the Oxford–Cambridge Arc. Computers, Environment and Urban Systems 83, 101515. https://doi.org/10.1016/j.compenvurbsys.2020.101515
Ovando, C., Pérez, J., Moral, A., 2015. LTE techno-economic assessment: The case of rural areas in Spain. Telecommunications Policy, New empirical approaches to telecommunications economics: Opportunities and challengesMobile phone data and geographic modelling 39, 269–283. https://doi.org/10.1016/j.telpol.2014.11.004
Owusu-Agyei, S., Okafor, G., Chijoke-Mgbame, A.M., Ohalehi, P., Hasan, F., 2020. Internet adoption and financial development in sub-Saharan Africa. Technological Forecasting and Social Change 161, 120293. https://doi.org/10.1016/j.techfore.2020.120293
Pan, S.J., Yang, Q., 2010. A Survey on Transfer Learning. IEEE Transactions on Knowledge and Data Engineering 22, 1345–1359. https://doi.org/10.1109/TKDE.2009.191
Parker, M., Acland, A., Armstrong, H.J., Bellingham, J.R., Bland, J., Bodmer, H.C., Burall, S., Castell, S., Chilvers, J., Cleevely, D.D., Cope, D., Costanzo, L., Dolan, J.A., Doubleday, R., Feng, W.Y., Godfray, H.C.J., Good, D.A., Grant, J., Green, N., Groen, A.J., Guilliams, T.T., Gupta, S., Hall, A.C., Heathfield, A., Hotopp, U., Kass, G., Leeder, T., Lickorish, F.A., Lueshi, L.M., Magee, C., Mata, T., McBride, T., McCarthy, N., Mercer, A., Neilson, R., Ouchikh, J., Oughton, E.J., Oxenham, D., Pallett, H., Palmer, J., Patmore, J., Petts, J., Pinkerton, J., Ploszek, R., Pratt, A., Rocks, S.A., Stansfield, N., Surkovic, E., Tyler, C.P., Watkinson, A.R., Wentworth, J., Willis, R., Wollner, P.K.A., Worts, K., Sutherland, W.J., 2014. Identifying the Science and Technology Dimensions of Emerging Public Policy Issues through Horizon Scanning. PLoS ONE 9, e96480. https://doi.org/10.1371/journal.pone.0096480
Peha, J.M., 2017. Cellular economies of scale and why disparities in spectrum holdings are detrimental. Telecommunications Policy 41, 792–801. https://doi.org/10.1016/j.telpol.2017.06.002
Perez, A., Yeh, C., Azzari, G., Burke, M., Lobell, D., Ermon, S., 2017. Poverty Prediction with Public Landsat 7 Satellite Imagery and Machine Learning. arXiv:1711.03654 [cs, stat].
Pokhriyal, N., Jacques, D.C., 2017. Combining disparate data sources for improved poverty prediction and mapping. PNAS 114, E9783–E9792. https://doi.org/10.1073/pnas.1700319114
Proville, J., Zavala-Araiza, D., Wagner, G., 2017. Night-time lights: A global, long term look at links to socio-economic trends. PLOS ONE 12, e0174610. https://doi.org/10.1371/journal.pone.0174610
PyTorch, 2020. PyTorch - An open source deep learning platform. [WWW Document]. URL https://www.pytorch.org (accessed 2.21.21).

Reddick, C.G., Enriquez, R., Harris, R.J., Sharma, B., 2020. Determinants of broadband access and affordability: An analysis of a community survey on the digital divide. Cities 106, 102904. https://doi.org/10.1016/j.cities.2020.102904

Reisdorf, B.C., Fernandez, L., Hampton, K.N., Shin, I., Dutton, W.H., 2020. Mobile Phones Will Not Eliminate Digital and Social Divides: How Variation in Internet Activities Mediates the Relationship Between Type of Internet Access and Local Social Capital in Detroit. Social Science Computer Review 0894439320909446. https://doi.org/10.1177/0894439320909446

Rhinesmith, C., Reisdorf, B., Bishop, M., 2019. The ability to pay for broadband. Communication Research and Practice 5, 121–138. https://doi.org/10.1080/22041451.2019.1601491

Riddlesden, D., Singleton, A.D., 2014. Broadband speed equity: A new digital divide? Applied Geography 52, 25–33. https://doi.org/10.1016/j.apgeog.2014.04.008

Righi, R., Samoili, S., López Cobo, M., Vázquez-Prada Baillet, M., Cardona, M., De Prato, G., 2020. The AI techno-economic complex System: Worldwide landscape, thematic subdomains and technological collaborations. Telecommunications Policy, Artificial intelligence, economy and society 44, 101943. https://doi.org/10.1016/j.telpol.2020.101943

Rosston, G.L., Wallsten, S., 2019. Increasing Low-Income Broadband Adoption through Private Incentives (SSRN Scholarly Paper No. ID 3431346). Social Science Research Network, Rochester, NY. https://doi.org/10.2139/ssrn.3431346

Rosston, G.L., Wallsten, S.J., 2020. Increasing low-income broadband adoption through private incentives. Telecommunications Policy 44, 102020. https://doi.org/10.1016/j.telpol.2020.102020

Sarkar, D., Bali, R., Ghosh, T., 2018. Hands-On Transfer Learning with Python: Implement advanced deep learning and neural network models using TensorFlow and Keras. Packt Publishing Ltd.

Saxe, S., MacAskill, K., 2019. Toward adaptive infrastructure: the role of existing infrastructure systems. Sustainable and Resilient Infrastructure 0, 1–4. https://doi.org/10.1080/23789689.2019.1681822

Schmid, T., Bruckschen, F., Salvati, N., Zbiranski, T., 2017. Constructing sociodemographic indicators for national statistical institutes by using mobile phone data: estimating literacy rates in Senegal. Journal of the Royal Statistical Society: Series A (Statistics in Society) 180, 1163–1190. https://doi.org/10.1111/rssa.12305

Sevastianov, L.A., Vasilyev, S.A., 2018. Telecommunication market model and optimal pricing scheme of 5G services, in: 2018 10th International Congress on Ultra Modern Telecommunications and Control Systems and Workshops (ICUMT). Presented at the 2018 10th International Congress on Ultra Modern Telecommunications and Control Systems and Workshops (ICUMT), pp. 1–6. https://doi.org/10.1109/ICUMT.2018.8631269

Simonyan, K., Zisserman, A., 2015. Very Deep Convolutional Networks for Large-Scale Image Recognition. arXiv:1409.1556 [cs].

Steele, J.E., Sundsøy, P.R., Pezzulo, C., Alegana, V.A., Bird, T.J., Blumenstock, J., Bjelland, J., Engø-Monsen, K., de Montjoye, Y.-A., Iqbal, A.M., Hadiuzzaman, K.N., Lu, X., Wetter, E., Tatem, A.J., Bengtsson, L., 2017. Mapping poverty using mobile phone and satellite data. Journal of The Royal Society Interface 14, 20160690. https://doi.org/10.1098/rsif.2016.0690

Steenbruggen, J., Tranos, E., Nijkamp, P., 2015. Data from mobile phone operators: A tool for smarter cities? Telecommunications Policy, New empirical approaches to telecommunications economics: Opportunities and challenges 39, 335–346. https://doi.org/10.1016/j.telpol.2014.04.001

Stevens, F.R., Gaughan, A.E., Linard, C., Tatem, A.J., 2015. Disaggregating Census Data for Population Mapping Using Random Forests with Remotely-Sensed and Ancillary Data. PLOS ONE 10, e0107042. https://doi.org/10.1371/journal.pone.0107042

Sultan, K., Ali, H., Zhang, Z., 2018. Call Detail Records Driven Anomaly Detection and Traffic Prediction in Mobile Cellular Networks. IEEE Access 6, 41728–41737. https://doi.org/10.1109/ACCESS.2018.2859756

Suryanegara, M., 2018. The Economics of 5G: Shifting from Revenue-per-User to Revenue-per-Machine, in: 2018 18th International Symposium on Communications and Information Technologies (ISCIT). Presented at the 2018 18th International Symposium on Communications and Information Technologies (ISCIT), pp. 191–194. https://doi.org/10.1109/ISCIT.2018.8588006

Tatem, A.J., 2017. WorldPop, open data for spatial demography. Sci Data 4, 1–4. https://doi.org/10.1038/sdata.2017.4

Taufique, A., Jaber, M., Imran, A., Dawy, Z., Yacoub, E., 2017. Planning Wireless Cellular Networks of Future: Outlook, Challenges and Opportunities. IEEE Access 5, 4821–4845. https://doi.org/10.1109/ACCESS.2017.2680318

Taylor, R.D., Schejter, A.M., 2013. Beyond Broadband Access: Developing Data-Based Information Policy Strategies. Fordham Univ Press.

Tchamyou, V.S., Erreygers, G., Cassimon, D., 2019. Inequality, ICT and financial access in Africa. Technological Forecasting and Social Change 139, 169–184. https://doi.org/10.1016/j.techfore.2018.11.004

The Africa Report, 2020. Ethiopia: 45% of telecoms company Ethio to be sold off, despite conflict in the north [WWW Document]. The Africa Report.com. URL https://www.theafricareport.com/51731/ethiopia-45-of-telecoms-company-ethio-to-be-sold-off-despite-conflict-in-the-north/ (accessed 1.8.21).

Thoung, C., Beaven, R., Zuo, C., Birkin, M., Tyler, P., Crawford-Brown, D., Oughton, E.J., Kelly, S., 2016. Future demand for infrastructure services, in: The Future of National Infrastructure: A System-of-Systems Approach. Cambridge University Press, Cambridge.

Turner-Lee, N.E., Miller, J.S., 2011. The Social Cost of Wireless Taxation: Wireless Taxation and its Consequences for Minorities and the Poor. Joint Center for Political & Economic Studies (Nov. 2011), available at http://www. jointcenter. org/sites/default/files/upload/research/files/The% 20Social% 20Cost% 20of% 20Wireless% 20T affixation. pdf ("The Social Cost of Wireless Taxation").

United Nations, 2019. The Sustainable Development Goals [WWW Document]. United Nations Sustainable Development. URL https://www.un.org/sustainabledevelopment/ (accessed 1.3.20).

Vesnic-Alujevic, L., Nascimento, S., Pólvora, A., 2020. Societal and ethical impacts of artificial intelligence: Critical notes on European policy frameworks. Telecommunications Policy, Artificial intelligence, economy and society 44, 101961. https://doi.org/10.1016/j.telpol.2020.101961

Vincenzi, M., Lopez-Aguilera, E., Garcia-Villegas, E., 2019. Maximizing Infrastructure Providers' Revenue Through Network Slicing in 5G. IEEE Access 7, 128283–128297. https://doi.org/10.1109/ACCESS.2019.2939935

Wallsten, S.J., 2001. An Econometric Analysis of Telecom Competition, Privatization, and Regulation in Africa and Latin America. The Journal of Industrial Economics 49, 1–19. https://doi.org/10.1111/1467-6451.00135

Wesolowski, A., Eagle, N., Noor, A.M., Snow, R.W., Buckee, C.O., 2012. Heterogeneous mobile phone ownership and usage patterns in Kenya. PloS one 7.

Whitacre, B., Strover, S., Gallardo, R., 2015. How much does broadband infrastructure matter? Decomposing the metro–non-metro adoption gap with the help of the National Broadband Map. Government Information Quarterly 32, 261–269. https://doi.org/10.1016/j.giq.2015.03.002

World Bank, 2021a. The World Bank In Malawi: Overview [WWW Document]. World Bank. URL https://www.worldbank.org/en/country/malawi/overview (accessed 1.8.21).

World Bank, 2021b. The World Bank In Ethiopia: Overview [WWW Document]. World Bank. URL https://www.worldbank.org/en/country/ethiopia/overview (accessed 1.8.21).

World Bank, 2019. Annual Report 2019 Lending Data. World Bank, Washington D.C.

World Bank, 2016a. Living Standards Measurement Study (LSMS) - Malawi 2016 [WWW Document]. URL https://microdata.worldbank.org/index.php/catalog/lsms (accessed 1.3.20).

World Bank, 2016b. Ethiopia Socioeconomic Survey 2015-2016 [WWW Document]. URL https://microdata.worldbank.org/index.php/catalog/lsms (accessed 1.3.20).

Yang, Q., Zhang, Y., Dai, W., Pan, S.J., 2020. Transfer Learning. Cambridge University Press, Cambridge. https://doi.org/10.1017/9781139061773

Zhou, Y., Smith, S.J., Zhao, K., Imhoff, M., Thomson, A., Bond-Lamberty, B., Asrar, G.R., Zhang, X., He, C., Elvidge, C.D., 2015. A global map of urban extent from nightlights. Environ. Res. Lett. 10, 054011. https://doi.org/10.1088/1748-9326/10/5/054011